\documentclass[
  journal=largetwo,
  manuscript=article-type,
  year=2020,
  volume=37,
]{cup-journal}

\usepackage{amsmath}
\usepackage[nopatch]{microtype}
\usepackage{booktabs,journal_abbr_mac}

\definecolor{ref@color}{RGB}{0,76,153}
\usepackage{hyperref}
\hypersetup{colorlinks=true,linkcolor=ref@color,citecolor=ref@color,urlcolor=ref@color}

\title{Dynamical aspects of Galactic habitability in N-body simulations}
 
\author{A. Mitra{\v s}inovi{\' c}}
\affiliation{Astronomical Observatory, Volgina 7, 11060 Belgrade, Serbia}
\email[A. Mitra{\v s}inovi{\' c}]{amitrasinovic@aob.rs}

\author{B. Vukoti\'c}
\affiliation{Astronomical Observatory, Volgina 7, 11060 Belgrade, Serbia}

\author{M. Micic}
\affiliation{Astronomical Observatory, Volgina 7, 11060 Belgrade, Serbia}

\author{M. M. \'Cirkovi\'c}
\affiliation{Astronomical Observatory, Volgina 7, 11060 Belgrade, Serbia}

 %% First letter not capped

\begin{document}

\begin{abstract}
Recent studies of Galactic evolution revealed that the dynamics of the stellar component might be one of the key factors when considering galactic habitability. We run an N-body simulation model of the Milky Way, which we evolve for 10 Gyr, to study the secular evolution of stellar orbits and the resulting galactic habitability-related properties, i.e., the density of the stellar component and close stellar encounters. The results indicate that radial migrations are not negligible, even in a simple axisymmetric model with mild levels of dynamical heating, and that the net outward diffusion of the stellar component can populate galactic outskirts with habitable systems. Habitable environment is also likely even at sub-Solar galactocentric radii, because the rate of close encounters should not significantly degrade habitability. Stars that evolve from non-circular to stable nearly-circular orbits typically migrate outwards, settling down in a broad Solar neighborhood. The region between $R \approx 3$ kpc and $R \approx 12$ kpc represents the zone of radial mixing, which can blur the boundaries of the Galactic Habitable Zone, as it has been conventionally understood. The present-day stable population of the stars in the Solar neighborhood originates from this radial mixing zone, with most of the stars coming from the inner regions. The Solar system can be considered as a typical Milky Way habitable system because it migrated outwards from the metal-rich inner regions of the Disk and has a circular orbit in the present epoch. We conclude that the boundaries of the Galactic Habitable Zone cannot be sharply confined for a given epoch because of the mixing caused by the stellar migrations and secular evolution of stellar orbits.
\end{abstract}

\section{Introduction}

In contemporary astrobiology, the habitability of galaxies is the most general overarching variable, defining the nature of astrobiological landscape on large spatiotemporal scales. The Galactic Habitable Zone (GHZ) was postulated as an extension and generalization of the Circumstellar Habitable Zone concept, initially just in the case of our Galaxy, the Milky Way \citep{Gonzalez+2001}, and subsequently to all kinds of galaxies \citep[e.g.][]{Suthar+McKay2012,2014MNRAS.440.2588S}. This zonal concept has formed a foundational view of galactic habitability and was initially metallicity-based with some dynamical input as well as temporal constraints on the age of potential habitats \citep{Lineweaver+2004}.

Metallicity is the basic galactic parameter that defines the amount of available building material for habitable planets \citep{Lineweaver2001}. Furthermore, metallicity, at least, roughly describes the availability of complex chemistry, which represents the ontological and evolutionary basis of life as we know it. Consequently, the inside-out model of metallicity buildup in disks of spiral galaxies \citep[e.g.][]{Schonrich+Binney2009,Frankel+2019,Johnson+2021}, has constrained the habitability estimates in a similar way. The GHZ was considered to be the annular ring that spreads outwards as the metallicity builds up in the outskirts of the galactic disk \citep{Lineweaver+2004}. \citet{Prantzos2008} argued that, at later epochs of the Milky Way history, the GHZ ring is more likely to expand than migrate outward, possibly making the whole disk suitable for hosting life. The metallicity-based considerations were also applied to the habitability of elliptical galaxies by \citet{Suthar+McKay2012}. These first models of galactic habitability have also considered the possibility of disruptive events, such as nearby gamma-ray bursts and supernovae, modeled through constraints on star formation rate. The stellar density was also factored in, to account for the total number of possible habitats. Considering the vertical distribution of stars in the Galactic disk as well, \citet{Gowanlock+2011} concluded that central regions of the Galaxy are the most habitable ones. 

However, the aforementioned studies lack the dynamical aspects of galaxy evolution (such as radial migration), which might impact the definition of the GHZ. Namely, the stars were considered to reside at nearly the same galactocentric radius during their lifetime without the ability to migrate to other parts of the Galactic disk. Moreover, they did not account for the possibility of spiral arms and Galactic plane crossings to increase the dangers from supernovae or Oort Cloud's objects. Several studies have investigated this possibility for the Solar system and concerning the Earth's mass extinction fossil record \citep{Clube+Napier1981,Rampino+Stothers1985,Raup+Sepkoski1986,Leitch+Vasisht1998,Goncharov+Orlov2003,Gillman+Erenler2008,Wickramasinghe+Napier2008,Filipovic+2013}, some of them significantly predating the GHZ studies. Although most of the habitability considerations from these works remain controversial \citep{Bailer-Jones2009,Feng+Bailer-Jones2013}, they implied that a robust assessment of galactic habitability should factor in the changes of the stellar environment along galactic orbits. Examining the orbits of more than 200 stars from the Solar neighborhood, \citet{PortodeMello+2014} stated that the Solar orbit is atypically circular, resulting in more spiral arms dwelling time than for the other examined stars, while \citet{Jimenez-Torres+2013} and \citet{BojnordiArbab+Rahvar2021} studied the effects of stellar fly-bys on the Solar system habitability, following the pioneer study of \citet{Laughlin+Adams2000}. 

Studies of Galactic evolution have partly attributed the observed radial metallicity gradient \citep[e.g., see ][]{Mayor1976,Spina+2021,Vickers+2021} to the possibility of radial stellar migrations \citep[for some of them see,][]{Sellwood&Binney2002,Haywood2008,Schonrich+Binney2009,Sanchez-Blazquez+2009,Minchev+Famaey2010,Bensby+2011,Lee+2011,Adibekyan+2013,Hayden+2015,Hayden+2018}. 
Interaction of stars with the disk structures, such as spirals or bars, cause the change in angular momentum, which results in radial (stellar) migration. Other possible migration causes might be the interaction with satellite galaxies \citep{Ruchti+2011,Bird+2012,Cheng+2012, Ramirez+2013} or stellar feedback \citep{El-Badry+2016}. \citet{Antoja+2018} specifically mention that the bar and spiral arms induce radial migrations (and that external perturbations from satellites also induce substructures). Furthermore, the authors argue that radial migrations might have influenced the MW disk to such an extent that it should not be considered to have an axial symmetry. \citet{Bird+2012} suggested that a signature of stellar migration could be the position of the Oort Cloud since it depends on the rate of stellar encounters (which correlates with stellar migrations). An extensive application of N-body (and hydrodynamical) simulations resulted in a better understanding of the stellar migrations, observed metallicity gradients and other galactic properties \citep{Roskar+2008,Minchev+2011,Brunetti+2011,Loebman+2011,Roskar+2012,Minchev+2012,Baba+2013,Kubryk+2013,Grand+2014,Loebman+2016}.

The mass resolutions of numerical simulations are at the order of a stellar cluster $(10^2-10^6\;\mathrm{M_\odot})$, with most of them having the stellar particle mass of $\sim10^4\;\mathrm{M_\odot}$. This has enabled a detailed understanding of individual galactic evolution and interactions within clusters of galaxies, and numerical simulations have become one of the most important tools of modern extragalactic astronomy. While we still lack sufficient computing power to resolve mass at the level of individual stars, numerical simulations can still prove useful for studying continuous habitability conditions in galaxies. They already made their way into the galactic habitability studies \citep{Vukotic+2016,Forgan+2017,Vukotic2017n,Stanway+2018,Stojkovic+2019MNRAS,Stojkovic+2019SerAJ}. It became evident that galactic habitability is far more subtle to understand than under the pioneering annular zone concept, which is still extensively used in circumstellar habitability. Galaxies are environments that are less centrally influenced in terms of radiation and movement when compared to the individual planetary systems and their host stars, so their habitability pattern should also appear different accordingly. They also have a 3-D structure and many other morphological features, such as spiral arms or bars, whose influence on habitability should not be neglected.

Contemporary understanding of the stellar migrations and advances made in the field of evolution of galaxies, as well as the nuanced nature of galactic habitability, have highlighted the need for investigating the dynamical aspects of habitability. More specifically, the aim of this work is to investigate the extent to which the stellar migrations can affect the very definition and boundaries of the GHZ, in a similar way they are used to describe and explain galactic metallicity patterns. To focus entirely on dynamical-related aspects of galactic habitability, pure N-body simulations of the Milky Way model are used to investigate only dynamical-related habitability constraints, contrary to the common practice thus far. As per \citet{Frankel+2020}, the different source causes of stellar migrations are separated as: notably "blurring" (driven by the radial heating) and "churning" (driven by the changes in the star's orbital angular momentum). The influence of these respective mechanisms on the boundaries of the GHZ is investigated.

Next section describes methods, models, and simulations and briefly analyze the dynamical evolution of the galaxy during $\sim10$ Gyr and the secular motions of stars. These aspects are further quantified by calculating actions. Afterward, this work considers possible habitability effects considering the rate of close encounters and orbital circularity given the radial migrations, particularly in the broad Solar neighborhood. Finally, after elaborating on the concept of the GHZ in light of the presented results, we give the conclusions of this work in the last section.

\section{Models and simulations}

We constructed two galaxy models using \texttt{GalactICs} software package \citep*{kuijken1995,widrow2005,widrow2008}. Both models consist of NFW \citep*{nfw1997} dark matter halo, exponential stellar disk, and \citet{hernquist1990} stellar bulge. Models have the same global physical parameters and differ only in particle resolution. For baryonic components, stellar disk, and bulge, we adopt the physical parameters of the commonly used MWb model \citep{widrow2005} as it satisfies observational constraints for the Milky Way. Hence, an exponential stellar disk with $3.53 \times 10^{10}$ M$_\odot$ total mass has a $2.817$ kpc scale radius and $0.439$ kpc scale height, while a stellar bulge with $1.51 \times 10^{10}$ M$_\odot$ total mass has $0.884$ kpc scale radius. In \texttt{GalactICs} software package, one also needs to model velocity structure for the disk component, through exponential radial dispersion profile\footnote{The authors of the software package state \citep{widrow2005}: \textit{"The dispersion in the azimuthal direction is related to [the radial one] through the epicycle equations, while the dispersion in the vertical direction is set by the vertical potential gradient and the vertical scale height."}} $\sigma_R^2(R)=\sigma_{R_0}^2 \textrm{exp}(-R/R_\sigma)$, where $\sigma_{R_0}$ is a central velocity dispersion (in our models, we adopt the value $124.4\; \mathrm{km\;s^{-1}}$) and $R_\sigma = 2.817\;\mathrm{kpc}$, scale radius is, for the sake of simplicity, equal to the spatial scale radius of the disk component. However, for the dark matter component, we adopt slightly different physical parameters: dark matter halo with $9.11 \times 10^{11}$ M$_\odot$ total mass has $13.16$ kpc scale, and concentration parameter $c=15$. This way, the model is a realistic representation of the Milky Way galaxy \citep[see][and references therein]{wang2020} with the total mass of $9.61 \times 10^{11}$ M$_\odot$. Moreover, the total mass enclosed within $100$ kpc, $M(R<100$ kpc$) = 7.41 \times 10^{11}$ M$_\odot$ is in line with recently reported observational constraints \citep[e.g.][]{CorreaMagnus+Vasiliev2022,Shen+2022ApJ...MW}. For the sake of simplicity, the dark matter halo and stellar bulge in our models do not rotate. Note that, while we list the final total masses of individual components in our models for practical reasons, the input parameters of the software package do not include the total mass\footnote{One can set the desirable total mass for the disk component only, but the resulting total mass of the disk can vary from the preset value if the software package fails to generate a stable disk with the required mass and structure.}, only structural and numerical parameters. The final total mass of individual components is obtained through the iterative procedure of stable model generation as the optimal one for a given set of the input parameters.

\begin{figure}
\includegraphics[width=\columnwidth,keepaspectratio=true]{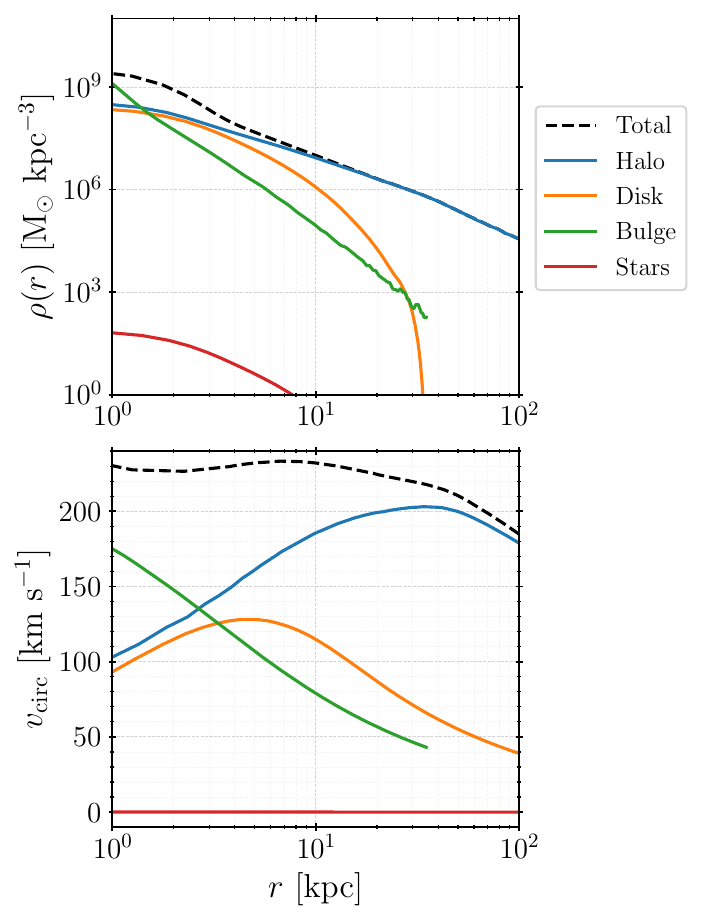}
\caption{Density profile (\emph{upper panel}) and idealized spherical rotational curve (\emph{lower panel}) of the galaxy model used with different line colors representing different subsystems, as indicated in the legend.}
\label{fig:galmodel}
\end{figure}

The higher resolution model, HRM, is composed of $2\times10^6$ particles in total, with $N_\mathrm{H} = 1\times10^6$, $N_\mathrm{D} = 7.12\times10^5$ and $N_\mathrm{B} = 2.88\times10^5$ particles in dark matter halo, stellar disk and stellar bulge, respectively. Lower resolution model, LRM, is scaled to have $4$ times less particles than HRM model, with $N_\mathrm{H} = 2.5\times10^5$, $N_\mathrm{D} = 1.78\times10^5$ and $N_\mathrm{B} = 0.72\times10^5$, resulting in total of $5\times10^5$ particles. Mass of a single baryonic particle is $\sim 2 \times 10^{5}$ M$_\odot$ in LRM, and $\sim 5 \times 10^{4}$ M$_\odot$ in HRM model. Additionally, we constructed another exponential stellar disk with the same spatial and velocity structure, consisting of $10^4$ particles with a total mass of $10^4$ M$_\odot$ where the mass of a single particle is equal to solar mass M$_\odot$. This  subsystem is referred to as "stars" and added to both models. Due to its non-zero total mass, it might appear as this artificially added subsystem will affect the evolution of the galaxy model. Density profiles and idealized spherical rotational curves of each subsystem, along with the total ones, are shown in Figure~\ref{fig:galmodel}. It is evident that the contributions from the stars subsystem are practically negligible, and the subsystem should not affect the evolution of the galaxy model in any meaningful way. Furthermore, the total mass of this subsystem is lower than the mass of a single baryonic particle in both galaxy models. The primary purpose of such an approach (adding the stars subsystem) is to ensure that the distribution is complete and that its spatial and velocity structure reflects that of a galactic disk. Analyzing the motions of all disk particles can be computationally costly and also tracking a set of random particles can introduce unwanted biases. Thus, given that this "stars" subsystem perfectly mimics the disk component of our model, its global long-term changes should be essentially the same as the changes we would observe analyzing the disk particles. It is also important to highlight that (since the added subsystem realistically represents the disk stars) individual stars are not all initially set on perfectly circular orbits, but rather the velocity structure of the subsystem dictates a certain distribution of orbital eccentricities expected in the galactic disk. As such, inner parts (e.g. $R<5\;\mathrm{kpc}$) host the majority of stars that are on highly eccentric orbits (i.e. non-circular), orbits of stars in the central region (e.g. $5\;\mathrm{kpc}<R<10\;\mathrm{kpc}$) are predominantly circular or nearly-circular, while the outskirts are populated exclusively with stars on nearly-circular orbits. This orbital distribution should not be considered as a limitation of the presented approach as the aim is to explore the dynamical effects on the secular evolution of individual stellar orbits in a realistic manner.

The models are evolved for $10$ Gyr using publicly available code \texttt{GADGET2} \citep{springel2000,springel2005gadget}, saving outputs (i.e. snapshots) every $0.1$ Gyr. The softening length parameter $\epsilon$, required in $N$-body simulations to limit the noise on small scales, in general, should take values that scale with the number of particles $N$ and dimensions of the system $R$ as $R/N^{1/2} < \epsilon < R/N^{1/3}$ \citep{b&t2008}. In practice, the optimal value for the softening length parameter remains somewhat ambiguous as several criteria have been proposed \citep[e.g.][]{merritt1996,dehnen2001,power2003,zhang2019}. We use the condition proposed by \citet{zhang2019}, $\epsilon = \alpha R/N^{1/2}$ with the value of free parameter $\alpha=2$, which yields the softening length of $\epsilon = 0.06$ kpc for baryonic particles in HRM, and $\epsilon = 0.12$ kpc in LRM model. As demonstrated by \citet{iannuzzi2013}, using a fixed value for softening length is a safe approach and does not affect the evolution of the disk component. At the same time, adopting a fixed value of softening length for all particle types (and thus using a sub-optimal value for dark matter particles) significantly reduces computational time.

\begin{table}
\begin{threeparttable}
\caption{List of simulations with relevant information: particle resolution $N$, adopted softening lengths and total computational time on $8$ CPU cores.}
\label{tab:sims}
\begin{tabular}{cccc}
\toprule
\headrow Simulation & $N$ & Softening & CPU time    \\
           &    ($10^6$)     & (kpc)     & (\texttt{h:m:s})     \\
\midrule
HRM        & 2.0       & 0.06     & \texttt{18:09:34.50}  \\ 
\midrule
LRM        & 0.5       & 0.12      & \texttt{04:20:46.42}  \\
\midrule
LRM\_UES   & 0.5       & 0.12\tnote{a}      & \texttt{32:32:35.63} \\
           &         & 0.80\tnote{b}       &             \\

\bottomrule
\end{tabular}
\begin{tablenotes}[hang]
\item[a]baryonic particles
\item[b]dark matter particles
\end{tablenotes}
\end{threeparttable}
\end{table}

However, the use of a fixed, constant softening length $\epsilon$ is justified only for higher resolution model \citep[as the resolution is comparable to that used by][]{iannuzzi2013}. A lower resolution model might still be (and probably is) sensitive to the choice of softening length parameter and its fine-tuning. To account for that, we run an additional simulation, named LRM\_UES, with unequal softening lengths for particles of different masses, where $\epsilon_\mathrm{DMP}=0.8$ kpc and $\epsilon_\mathrm{BP}=0.12$ kpc are optimal softening lengths for dark matter and baryonic particles, respectively.

In addition to limiting the noise on small scales, softening length $\epsilon$ regulates the integration timestep $\Delta t$ in \texttt{GADGET2}: lowering softening length lowers the minimum integration timestep. To compare the results of different simulations, we thus need to keep the minimum integration timestep constant in all runs, which we achieve by varying the accuracy of time integration \citep[parameter \texttt{ErrTolIntAccuracy},][]{springel2000,springel2005gadget}. Relevant information for all simulations (names, particle resolution, adopted softening lengths), as well as total computational times (CPU time) using $8$ CPU cores with $3.50$ GHz base frequency and $16$ MB cache, is summarized in Table~\ref{tab:sims}. 

Since this work requires a stable and robust model, the total energy, and angular momentum must be conserved during the evolution, which is satisfied in all simulations. We also require that relevant radial profiles (namely, density and dispersion of velocity components) do not change significantly. Some minor changes are not only allowed but expected, as they will inevitably occur in numerical models due to dynamical heating \citep[e.g.][]{Sellwood2013}. Moreover, since we did not implement any internal perturbations or instabilities, we also require that models remain axisymmetric (i.e., that the bar or transient spiral structure does not emerge). All of these requirements are met in simulations LRM\_UES and HRM, while, in LRM, the dynamical heating is pronounced, and the model does not retain axial symmetry. As a consequence, its relevant radial profiles change significantly. We will briefly discuss this in the following section.

\section{Galaxy evolution and the motions of stars}

\begin{figure*}
\includegraphics[width=0.95\textwidth,keepaspectratio=true]{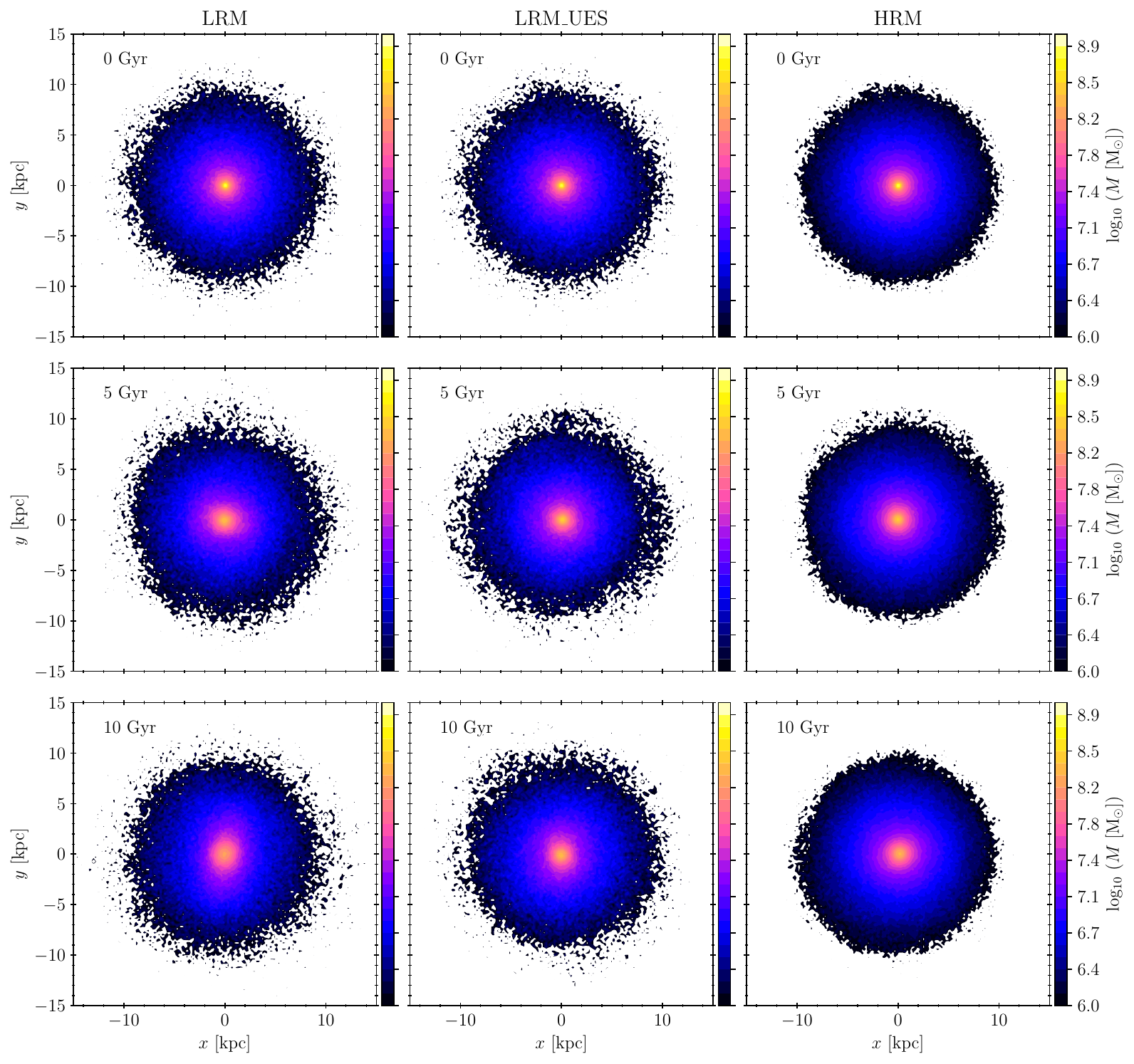}
\caption{Mass distribution of the stellar galaxy components (disc+bulge) at three different times (\emph{top to bottom:} $t\in\{0,5,10\}$ Gyr) for all simulations (\emph{left to right:} LRM, LRM\_UES, HRM) in face-on projection, i.e. in $x-y$ plane.}
\label{fig:gal-xy}
\end{figure*}

\begin{figure*}
\includegraphics[width=0.95\textwidth,keepaspectratio=true]{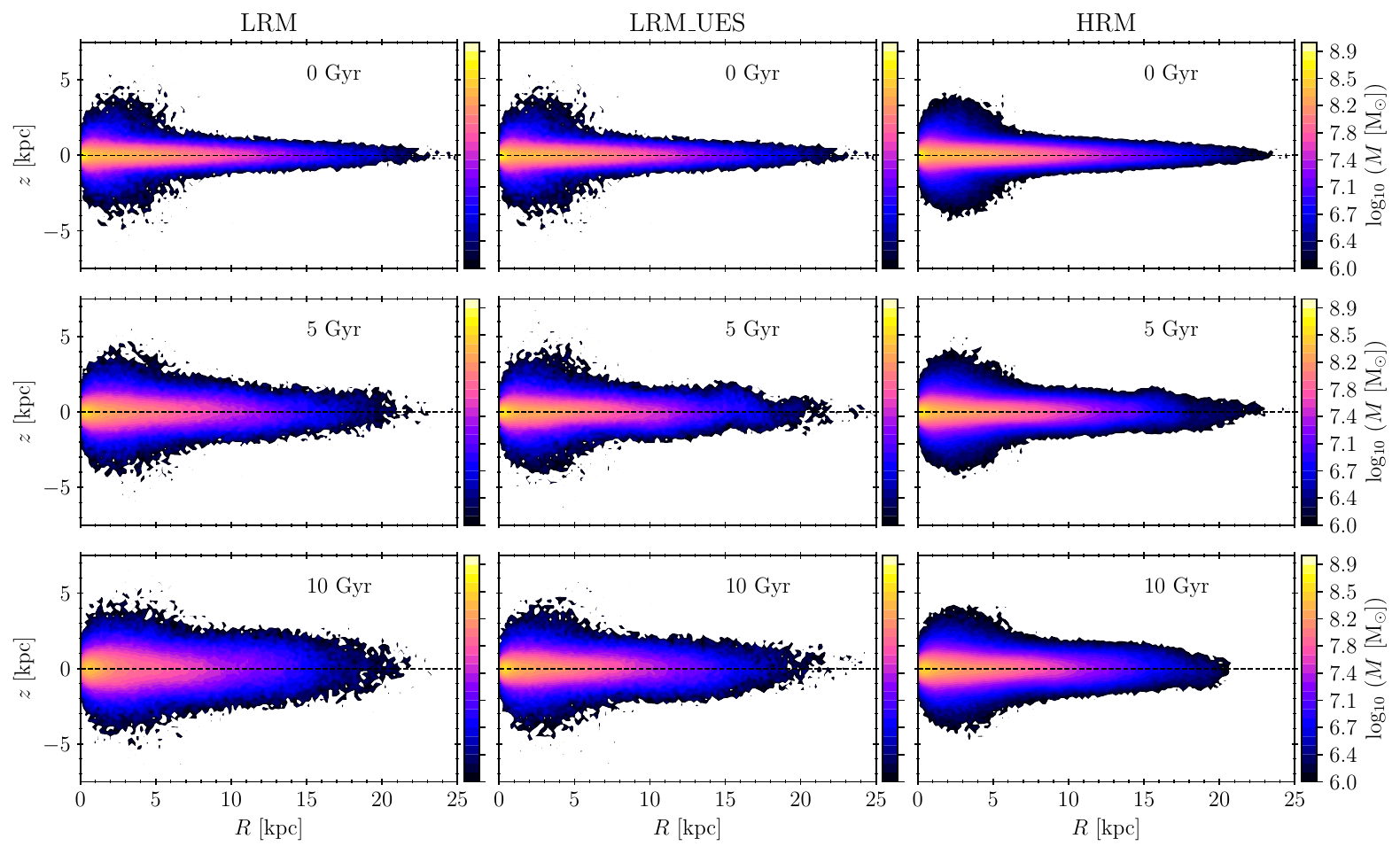}
\caption{Same as Figure~\ref{fig:gal-xy}, in $R-z$ plane.}
\label{fig:gal-rz}
\end{figure*}

We start our analysis by centering the galaxy model on the stellar (disc+bulge) center of mass and rotating the model so that the angular momentum of the disk is aligned with the positive direction of $z$-axis in order to ensure that the disk plane is positioned in the $x-y$ plane and its rotation is direct. Mass distributions of the stellar galaxy components, for all simulations, at three different times ($t\in\{0,5,10\}$ Gyr) are shown in Figure~\ref{fig:gal-xy} (face-on projection, $x-y$ plane) and Figure~\ref{fig:gal-rz} ($R-z$ plane). By the end of the simulation, a moderately strong bar forms in LRM, while the galaxy model in both LRM\_UES and HRM remains axisymmetric. Bar formation is affected by the angular momentum transfer between the disk and dark matter halo particles. Thus, using a live halo is preferred over static potential, and particle resolution plays a significant role. While \citet{Weinberg+Katz2007} argued that the dark matter halos need to be resolved with more than $10^8$ particles to minimize numerical noise, \citet{Sellwood2008} deemed this excessive, suggesting that $10^6$ particles are sufficient. Only our HRM simulation satisfies this condition, but we can also consider the data generated with LRM\_UES.

Dynamical heating of the system and disk thickening is the most pronounced in LRM simulation, as seen in Figure~\ref{fig:gal-rz}, while the differences between LRM\_UES and HRM appear marginal. The significant dynamical heating in the lower-resolution model is expected and in good agreement with the findings of \citet{Sellwood2013}. Interestingly, our results imply that adopting optimal softening lengths for different particle types can negate this effect and that the significance of the softening length inversely correlates with particle resolution. Careful adoption of appropriate softening lengths is thus crucial to avoid artificial numerical effects and consequences on lower particle resolutions. However, taking into account the computational resources required for such an endeavor and the total CPU time, compared with the higher particle resolution case (Table~\ref{tab:sims}), it should be evident that such a solution is not optimal and that opting for higher resolution models should be preferred.

\begin{figure}
\includegraphics[width=\columnwidth,keepaspectratio=true]{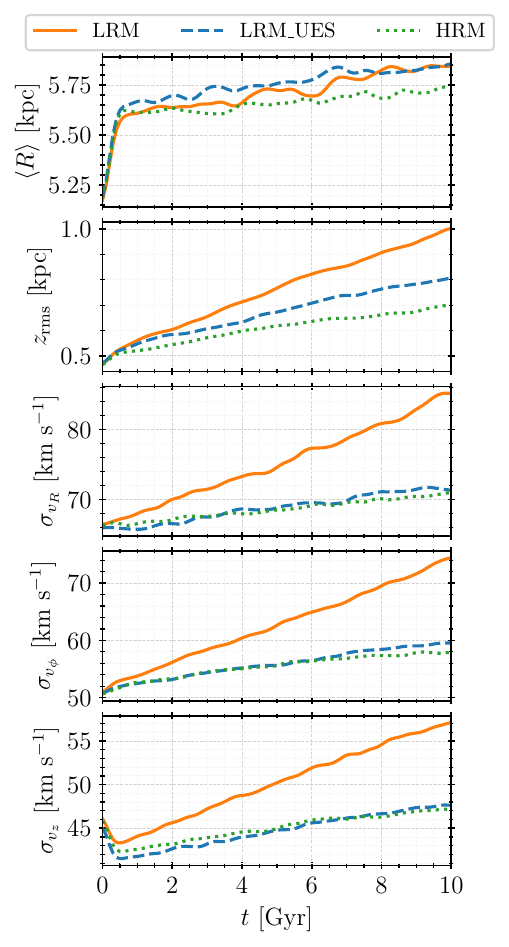}
\caption{Evolution of the global properties of the stars subsystem, top to bottom: median galactocentric distance $\langle R \rangle$, (arithmetic) mean height $z_\mathrm{rms}$ and velocity dispersion in cylindrical coordinates: radial $\sigma_{v_R}$, circular $\sigma_{v_\phi}$ and vertical $\sigma_{v_z}$. Different simulations are represented with different line colors and styles, as indicated by the legend.}
\label{fig:starsbasic}
\end{figure}

The evolution of basic parameters of the stars subsystem should indicate differences between the simulations in a more transparent way. We show it in Figure~\ref{fig:starsbasic}. These basic parameters include median galactocentric distance $\langle R \rangle$, (arithmetic) mean height $z_\mathrm{rms}$ and velocity dispersion in cylindrical coordinates: radial $\sigma_{v_R}$, circular $\sigma_{v_\phi}$ and vertical $\sigma_{v_z}$. The artificially added stars subsystem needs about $0.5$ Gyr to stabilize in galaxy models - consequently, we will use the snapshot corresponding to $t = 0.5$ Gyr as an initial one. Median galactocentric distance $\langle R \rangle$ slowly increases, slightly more so in simulations with lower particle resolutions, but the difference is negligible on a larger scale. It implies that the stars subsystem, as a whole, has a minor shift outwards in all simulations, meaning that the outward radial migration is slightly more substantial than its inward counterpart \citep[in agreement with][]{Roskar+2012}. Individual stars can still experience drastic radial migrations, but the ones migrating inwards mostly balance out the ones migrating outwards.

Mean heights $z_\mathrm{rms}$, indicative of disk thickness, support our previous observations. Namely, in LRM simulation, the disk practically doubles its thickness, primarily due to bar formation. The significant increase in individual velocity dispersion indicates this in particular, as the velocity dispersion is higher in barred galaxies, compared to their non-barred counterparts \citep[e.g.][]{Kormendy1983, Bettoni+1988}. While the disk thickness increases more in LRM\_UES, compared to HRM simulation, the evolution of their respective velocity dispersion is practically indistinguishable. In both simulations, velocity dispersion also increase (although at a much slower rate than in the LRM simulation), which is a sign of the dynamical heating of the system. However, the absolute magnitude of the said increase, considering it spans over the entire course of simulations (i.e. over $10$ Gyr), is minor on a global scale.

For the remainder of this work, i.e. for detailed analysis of the motions and migrations of stars, as well as habitability considerations, we will focus on HRM simulation only. Preliminary analysis presented here indicate that there should be no significant differences between HRM and LRM\_UES simulations. Thus, the results of the detailed analysis should appear roughly the same, accordingly. The other lower-resolution simulation, LRM, does not satisfy our stability criteria as the dynamical heating appears extreme, and the model deviates from the axial symmetry. These effects are purely numerical, arising due to low particle resolution coupled with the adoption of improper softening lengths, and, as such, the bar formation is artificial and non-realistic. Star particles in this simulation are prone to the same numerical effects, and we cannot consider the results of the detailed analysis of their motions reliable. However, for the sake of completeness, we will briefly analyze radial migrations in all models and make a comparison with our main results.

\subsection{Action calculation and analysis}

Once the model of the galaxy is appropriately centered and rotated, we utilize \texttt{AGAMA} \citep{Vasiliev2019}, a publicly available software library for a broad range of applications in the field of stellar dynamics. A smooth approximation of the galactic potential is generated and used to calculate three standard action coordinates for star particles. These are the vertical action $J_z$, the radial action $J_R$, and the azimuthal action $J_\phi$. Combined, they fully describe the orbit of the star particle: $J_R$ and $J_z$ describe oscillations in the radial and vertical directions, respectively, while the azimuthal action represents the $z$-component of the angular momentum $L_z$ in an axisymmetric potential. We will adopt the notation of $L_z$ for the azimuthal action for the remainder of this work.

Additionally, the angular momentum of the particle $L_\mathrm{c}(E)$ is calculated with total energy $E$, for a circular orbit \citep[e.g.][]{Abadi2003}. Then, the circularity of the orbit can be calculated as $\xi = L_z/L_\mathrm{c}(E)$, for each star particle. Circularity, defined this way, is just another way of assessing the eccentricity of the orbit. For example, particles with $\xi \geq 0.9$ are on nearly circular orbits, while the ones with lower circularity values have more eccentric orbits.

\begin{figure*}
\includegraphics[width=0.98\textwidth,keepaspectratio=true]{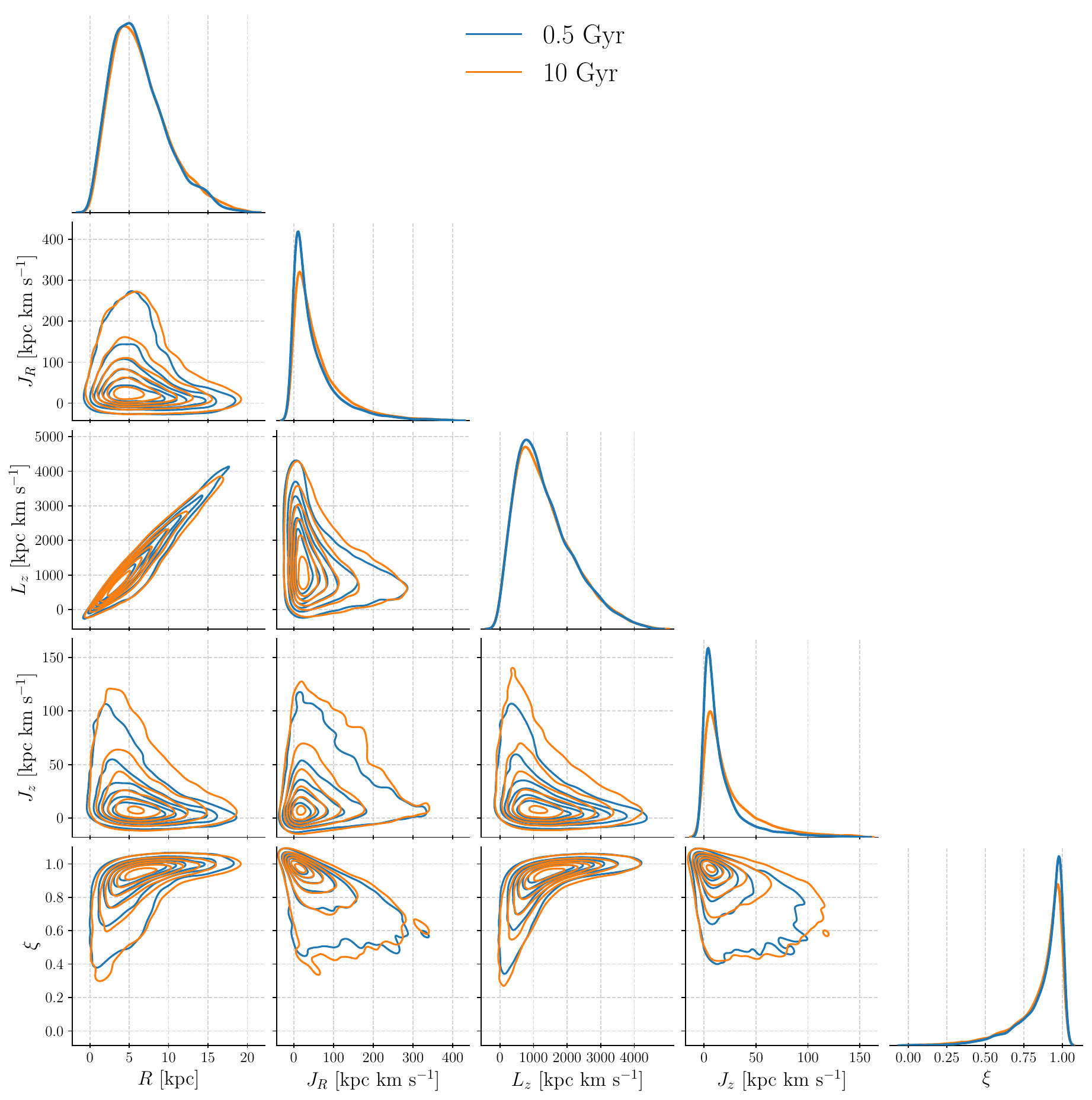}
\caption{Probability density distributions, $R$ is galactocentric distance, $J_R$, $L_z$ and $J_z$ are radial, azimuthal and vertical actions, respectively, and $\xi$ is orbital circularity, for two different times represented with different line colors, as indicated by the legend.}
\label{fig:cornerdist}
\end{figure*}

Figure~\ref{fig:cornerdist} shows the distributions of these parameters - actions, circularity, and galactocentric distance, at two different times (initial and final). These distributions do not change significantly throughout the simulation, which is expected in a stable axisymmetric model of the Milky Way. Minor disk thickening is noticeable on distributions that include vertical action $J_z$. The apparent immutability of distributions, a sign of the stability of the system as a whole, does not imply that individual stars do not migrate. It merely indicates that most migrations are balanced out on a larger scale. As the migrations of individual star particles are of the most interest for this work, we will focus on them once we state key observations from Figure~\ref{fig:cornerdist}. The majority of particles are in highly circular orbits, as is both intuitively expected and observationally grounded, as far as the Milky Way stellar population is concerned \citep[e.g.][]{cubarsi2021orbital}. The ones whose orbits are significantly eccentric (with lower circularity values) inhabit, almost exclusively, inner to central regions of the galaxy. This includes a broad Solar neighborhood, in its entirety. Hence, even without looking into individual migrations, it is expected that the motions of stars affect the boundaries of the GHZ, at least to some extent. Stars that are not on nearly circular orbits \citep[stars with circularities $\xi<0.9$, e.g.][]{BeraldoeSilva+2021}, account for $39.5\%$ of our stars subsystem, and the majority of them are confined within inner $9.65$ kpc initially (at $t=0.5$  Gyr), while the values change to $45.3\%$ and $10.17$ kpc by the end of the simulation (at $t=10$ Gyr). These stars, as expected, have lower values of angular momentum $L_z$, while both their radial $J_R$ and vertical $J_z$ action span over a broad range. Interestingly, the $J_R-J_z$ distribution in Figure~\ref{fig:cornerdist}, shows explicit anti-correlation: stars with higher radial action have lower values of vertical counterpart and vice versa. This means that stars whose orbits deviate from nearly circular tend to oscillate either in a radial or in a vertical direction, but not simultaneously in both of these directions if the oscillations are drastic.

\begin{figure}
\includegraphics[width=0.85\columnwidth,keepaspectratio=true]{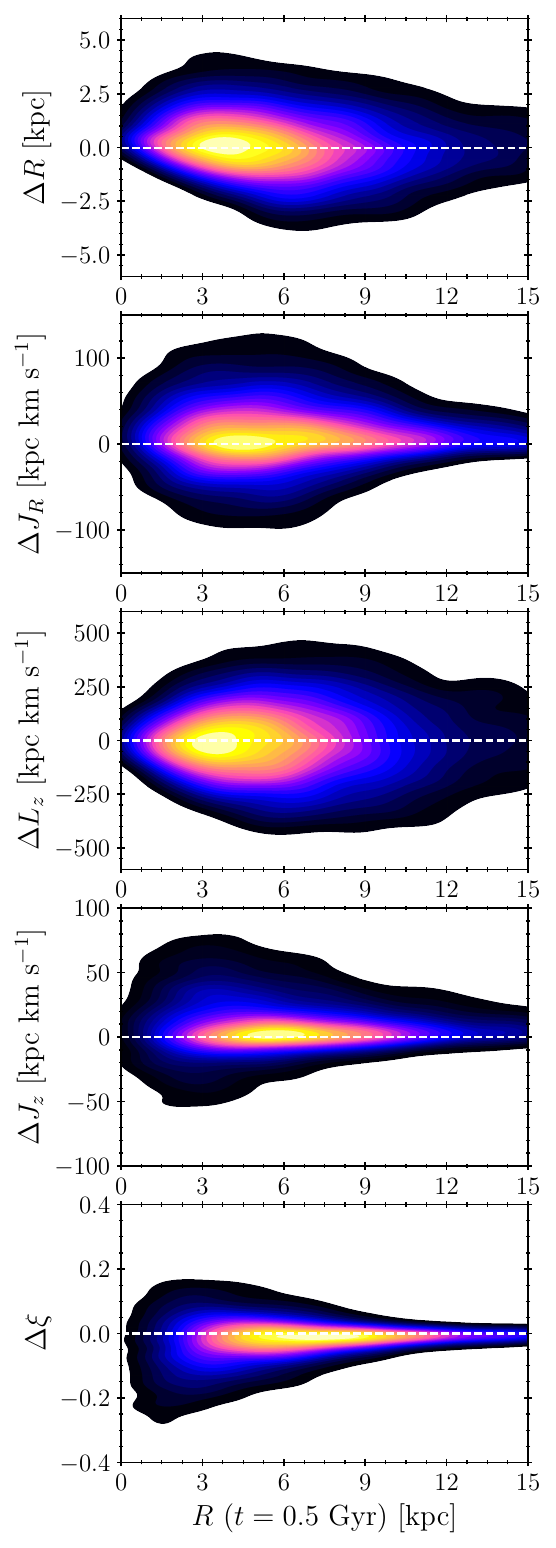}
\caption{Two-dimensional probability density distributions where $x$-axis, $R\;(t=0.5\;\mathrm{Gyr})$, is the initial galactocentric distance and $y$-axis corresponds to the absolute change of parameters defined and presented in Figure~\ref{fig:cornerdist}. The absolute change is calculated as: $\Delta X = X(t=10\;\mathrm{Gyr})-X(t=0.5\;\mathrm{Gyr})$, where $X$ is any of the parameters.}
\label{fig:deltas}
\end{figure}

For individual stars, we calculate the absolute changes of relevant parameters (galactocentric distance $R$, circularity $\xi$ and actions $J_R$, $L_z$ and $J_z$) as:
\begin{equation}\label{eq:delta}
 \Delta X = X(t=10\;\mathrm{Gyr})-X(t=0.5\;\mathrm{Gyr})   
\end{equation}
\noindent where $X$ is any of the parameters. We show two-dimensional probability density distributions $R\;(t=0.5\;\mathrm{Gyr})-\Delta X$, where $R\;(t=0.5\;\mathrm{Gyr})$ is an initial galactocentric distance, in Figure~\ref{fig:deltas} to determine the magnitude of these changes and which regions of the disk are the most prone to them.

Angular momentum change, $\Delta L_z$, is the only considered parameter whose distribution appears almost perfectly symmetrical around zero. In total, $51.2\%$ of stars satisfy $\Delta L_z<0$, i.e. they experience angular momentum loss (it is implied that the rest have angular momentum gain). The most extreme angular momentum absolute changes (of any sign) correspond to stars initially located in the central region. Similarly, the most extreme absolute changes in the radial action $J_R$ correspond to stars initially located in inner to central regions, although the probability density distribution is not as symmetric around zero, and more stars experience positive change (with only $37.8\%$ of the sample satisfying $\Delta J_R<0$). The majority of stars are evolving toward orbits that oscillate more radially. On the contrary, both vertical action $J_z$ and circularity $\xi$ have the most extreme absolute changes in the innermost regions, and both probability density distributions are asymmetrical around zero. Across all galactocentric distances, vertical action predominantly increases (in total, $67.3\%$ of stars satisfy $\Delta J_z>0$), which is in line with previously discussed disk thickening. Possibly as a consequence, circularity predominantly decreases (in total, $68\%$ of stars satisfy $\Delta \xi<0$). However, at higher distances, the absolute changes in circularity are minor, indicating that most stars that inhabit the outer regions do not have significant deviations from their initial circularity or orbital eccentricity. Interestingly, there are stars whose circularity increases. However small a fraction it is, this sub-sample of stars appears to be the most interesting from the habitability point of view: these stars evolve toward circular orbits, which means that their galactic environment conditions are likely to be more stable in the long term. We will consider this sub-sample in particular when analyzing and discussing habitability.

Finally, the parameter of the most interest, the galactocentric distance, has the probability density distribution of its absolute change asymmetrical around zero. In total, about $46\%$ of stars migrate inwards (i.e. satisfy $\Delta R<0$), while the rest migrate outwards (i.e. $\Delta R>0$), resulting in the net radial migration outwards of about $8\%$\footnote{To validate that the stars subsystem indeed mimics the galactic disk, we calculated radial migration rates of disk particles and found the same net outwards trend of about $8\%$.}. This result is in line with the report of e.g. \citet{Roskar+2008,Roskar+2012}, who noticed that the outward radial migration is larger than its inward counterpart and our previous observation that the minor global radial migration outward is expected. The asymmetry of this probability density distribution is insightful. Interestingly, stars with the most extreme outward radial migrations are initially clustered around $R\simeq 4.44\;\mathrm{kpc}$, while the ones with significant inward migrations originate from higher galactocentric distances, $R\simeq 6.84\;\mathrm{kpc}$. This hints at a sort of radial mixing in the central disk parts and should, in principle, affect the traditionally constrained GHZ.

It is important to highlight that the radial mixing is typically associated with spiral patterns \citep[e.g.][]{Sellwood&Binney2002} and that the previous studies, mentioned earlier, include transient spiral structure. The general agreement of our results implies that the spiral patterns are not a necessary condition for radial migrations and mixing and that similar outcomes can happen in axisymmetric models with sufficient, mild levels of dynamical heating. However, the inclusion of a transient spiral structure in our model would, most likely, result in slower radial migration outwards or a lower net outwards trend at the expense of even more efficient radial mixing in the central parts of the disk.

As mentioned in the introductory section, the aim is to separate different source causes of stellar radial migrations. For this purpose, it is investigated if there is a correlation between the relative changes in actions and the radial changes and of what significance and strength. More specifically, the Spearman's correlation coefficient $\rho$ is calculated between $\Delta X/X_0$ and $\Delta R$, where $X$ corresponds to any of the three actions ($J_R$, $L_z$ or $J_z$), $\Delta X$ is defined with Equation~\ref{eq:delta} and $X_0$ represents appropriate initial value at $t=0.5\;\mathrm{Gyr}$. The results of this test are listed in Table~\ref{tab:spearmanr}. Since the test is sensitive to outliers, a robust linear regression is performed (thus, de-weighting outliers) with $95\%$ confidence interval. No difference is found, and both tests lead to essentially the same conclusions.

\begin{table}
\begin{threeparttable}
\caption{Spearman's correlation coefficient $\rho$ between relative change $\Delta X/X_0$ and absolute change $\Delta R$ in the galactocentric distance, where $X$ corresponds to any of the three actions, $\Delta X$ is defined with Equation~\ref{eq:delta} and $X_0$ represents initial value at $t=0.5\;\mathrm{Gyr}$.}
\label{tab:spearmanr}
\begin{tabular}{ccc}
\toprule
\headrow Action & Correlation $\rho$ & $p$-value    \\
\midrule 
$J_R$        & $+0.0956$       & $1.32 \cdot 10^{-20}$  \\
\midrule
$L_z$        & $+0.4034$       & $0.0000$  \\
\midrule
$J_z$        & $-0.0128$       & $0.2127$     \\

\bottomrule
\end{tabular}
\end{threeparttable}
\end{table}

Relative changes in angular momentum and the radial action are correlated with the radial change, but the relative change in the vertical action appears independent. In particular, the correlation is moderate for angular momentum and very weak for the radial action. This is expected and corroborates the conclusion of \citet{Frankel+2020} that diffusion in angular momentum dominates in the secular orbit evolution and that, in comparison, radial heating is much weaker.

Vertical motions appear stochastic and show no explicit correlation with radial migrations. They might be irrelevant to the present study, as its focus is on the radial boundaries of the GHZ. This is manifestly not the case, however, when the habitability of disk galaxies is considered on a larger scale. Vertical motions directly affect the number of galactic plane crossings, which can be hazardous as the probability of close encounters peaks in mid-plane \citep[e.g.][]{Medvedev+Melott2007,Bailer-Jones2009,Melott+2012,Sloan+2017}, as do tidal stresses associated with galactic disk. Moreover, stellar orbits with higher vertical oscillations contribute to these stellar systems experiencing vastly different environments, which might not satisfy, in terms of stability, continuous habitability conditions. Any attempt to unify various habitability conditions and constraints and consider multiple factors toward a complex galactic habitability model should, thus, include vertical motions in some way. In such a case, it is essential to separate kinematically distinct (among other, non-dynamical differences) thin and thick disk components \citep[see, e.g.][and references therein]{Vieira+2022}, as their highest impact on the results should be related to the vertical motions. The single-disk model from this work, which should not be solely considered as representative of a thin disk, is suitable for this study, which aims to make a first step toward exploring and quantifying dynamical-related effects on the radial boundaries of the GHZ. However, when vertical motions are taken into consideration, the single-disk model becomes less applicable as multiple-component disk models should uncover insightful trends.

\subsection{Comparison with lower-resolution models}

For the sake of completeness, as previously mentioned, radial migration trends are briefly analyzed for all models and compared with the main results. As expected, in the LRM\_UES simulation, radial migration trends are roughly the same, with a net outwards trend of roughly $8\%$, and the asymmetry of $R-\Delta R$ distribution around zero being the same. This could confirm the previous remark that lower-resolution models could still give reliable results if numerical parameters, such as softening lengths of different particles, are carefully chosen.

On the contrary, simulation LRM considerably differs from the main results. The whole disk expands significantly in the vertical direction, and a net outwards trend of radial migrations is slightly higher (around $10\%$). The absolute magnitudes of radial migrations, i.e. $|\Delta R|$ values, are typically larger than in our main simulation, suggesting that the bar might be able to induce migrations of a longer range. Stars in the inner region of the disk, where the artificially-formed bar is located, predominantly migrated inwards (around $62\%$ of stars satisfy $\Delta R < 0$). Median magnitudes of radial migrations in this region are also angle-dependent: the largest migrations outwards are aligned with the bar's major axis, whereas their inwards counterparts align with the minor axis. Unsurprisingly, as the stars captured by the bar evolve toward radial orbits, the negative changes of circularity are, on average, higher in magnitude than in our main simulation and aligned with the bar's major axis. In the parts of the disk toward the central region, however, we did not notice signs of efficient radial mixing, as the $R-\Delta R$ distribution appears roughly symmetrical around zero (contrary to LRM\_UES and HRM simulations).

While the results of the LRM simulation are not considered to be reliable, due to numerical effects, there is a striking agreement between the brief results presented here and the previous more rigorous works on the topic \citep[see, e.g.][and references therein]{DiMatteo+2013,Filion+2023}. The lack of efficient radial mixing in the presence of a bar should have been expected. As \citet{DiMatteo+2013} point out, in non-axisymmetric galaxy models, migrations and mixing do not occur at the same time, and radial mixing can only be established when the phase of significant radial migrations, caused by the bar, is over. Angle-dependent trends of radial migrations and the inwards trend in the inner disk region are also in agreement with the results of \citet{Filion+2023}. However, despite these similarities, we cannot use this model in our habitability considerations due to the previously mentioned numerical effects that contaminate the results. For the results to be reliable and robust enough, our model has to have numerical effects as low as possible and the bar formation should arise from an actual (realistic) instability.

\section{Habitability considerations}

The primary simplifications of this study are the lack of gas and, consequently, the lack of star-formation rates and metallicity information. In a way, this limits possible habitability-related considerations, in particular the defining of the GHZ in its traditional sense. However, it is certainly not impossible to discuss habitability in multiple ways. Since the dynamical issues have been mentioned but mostly skirted around, ever since the emergence of the GHZ concept, it makes sense to use the well-developed apparatus of N-body simulations to clearly separate dynamical parameters from the rest of the parameters of galactic habitability \citep[for extensive discussion of those see,][]{Stojkovic+2019SerAJ}. 

In what follows, the rate of close encounters is calculated and discussed, which is a typical, strictly dynamical constraint on the GHZ and habitability in general. We will also briefly analyze a subset of stars evolving from non-circular to nearly circular orbits, as these are particularly interesting from an astrobiological point of view. Finally, since the galaxy model in this work corresponds to the Milky Way, the broad Solar neighborhood will be closely examined, and also the effect of radial migrations on this ring-shaped zone and its stellar population.

\subsection{The rate of close encounters}
\begin{figure}
\includegraphics[width=\columnwidth,keepaspectratio=true]{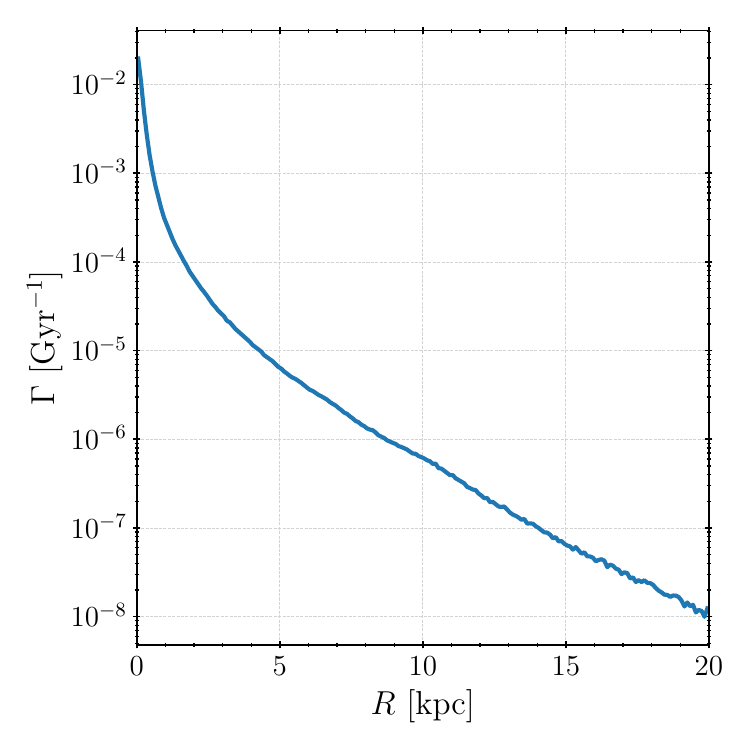}
\caption{The rate of close encounters $\Gamma$, defined with Equation~\ref{eq:gamma}, as a function of galactocentric distance $R$.}
\label{fig:rateenc}
\end{figure}

Aside from the metallicity condition, which defines the availability of material to form complex forms of life, another critical condition for habitability is the continuity of habitable-friendly conditions \citep[][for example, call this notion "habitable time"]{Vukotic+2016}. From a strictly dynamical point of view, this continuity will be disrupted by the close stellar encounters, which are catastrophic enough to change planetary orbits initially in circumstellar habitable zones. The rate of such encounters $\Gamma$ is given by:
\begin{equation}\label{eq:gamma}
\Gamma =  n_\star \langle v \rangle \langle \sigma \rangle
\end{equation}
\noindent where $n_\star$ is stellar number density, $\langle v \rangle$ relative stellar velocity and $\langle \sigma \rangle$ cross-section of the encounter having adverse astrobiological consequences. Evidently, we can directly calculate $\langle v \rangle$, and we will adopt $\langle \sigma \rangle \sim 100\;\mathrm{AU}^2$ \citep[as suggested by, e.g.][]{Laughlin+Adams2000}. Stellar number density estimate is particularly challenging since we can only calculate mass density $\rho_\star$ in our galaxy model and do not have information on the number of stars. For the sake of simplicity, we will assume that all stars have Solar masses, i.e. that $n_\star = \rho_\star/\mathrm{M}_\odot$. Since our galaxy model is axisymmetric, it is safe to calculate the rate of close encounters $\Gamma$ as a function of galactocentric distance $R$, which we show in Figure~\ref{fig:rateenc}.

Represented this way, the rate of close encounters is a declining function of galactocentric distance. This means that imposing a certain threshold to limit the boundaries of the GHZ is only possible for the inner one. The outer boundary is, rather inconveniently for this study, limited by the metallicity. Generally speaking, one can argue that the outer boundary of the GHZ is also limited by the environment where the galaxy resides (e.g. part of a cluster or group, having satellite galaxies, etc.), given the outside-in nature of tidal effects. High-resolution cosmological or zoom-in simulations should be the best currently available tool for studying these environmental effects and limiting the outer boundary of the GHZ based on those considerations, in addition to metallicity.

To constrain the inner boundary of the GHZ, a critical value for the rate of close encounters needs to be assumed, for example, $\Gamma_\mathrm{crit} = 1 \;(\mathrm{age\;of\;the\;Earth})^{-1} \simeq 0.22\;\mathrm{Gyr}^{-1}$. The calculated rate of encounters is smaller than this critical value over the entire range, which implies that the inner GHZ boundary cannot be constrained in this manner. Apart from the relevance of the inner boundary determination, the precision of the estimate can also be scrutinized. We will address the latter part first. For the Solar system, $\Gamma (\mathrm{R_\odot}) \sim 10^{-6}\;\mathrm{Gyr}^{-1}$, but \citet{Sloan+2017} give an estimate of $3 \times 10^{-8}\;\mathrm{Gyr}^{-1}$ for this value (thus, even smaller). While rigorously studying the effects of stellar encounters on habitability, \citet{BojnordiArbab+Rahvar2021} found the expected number of threatening stellar encounters in the Solar neighborhood is $\Gamma \sim 10^{-4}\;\mathrm{Gyr}^{-1}$. Thus, the estimate from this work, falls exactly in the middle and is in line with previous findings. Despite this agreement, our estimate might be too crude for the inner parts of the disk. More specifically, our assumption for the stellar number density is most likely incorrect in these regions, as the stars may have, on average, lower mass than the assumed Solar mass. Accounting for this may increase the rate of close encounters by an order of magnitude in the central region, which is still not enough of an increase to constrain the GHZ outside the central galactic region. Another argument that can be made is that our adopted value for cross-section is too conservative and that higher values should be used \citep[see, e.g.][for cross-section discussion]{Li+Adams2015,Brown+Rein2022}. The closest encounters of the Solar system within the present $10^7$ yr are found to occur at a distance of $10^4$ AU from Gaia Data Release 3 \citep{2022ApJ...935L...9B}. As adopting this value for the cross-section in our calculation might imply an overly restrictive view of habitability, we will consider its implications in relation to previously presented results. Since the cross-section is de facto a constant, adopting $\langle \sigma \rangle \sim 10^4\;\mathrm{AU}^2$ instead would not change the shape of the declining $\Gamma (R)$ function shown in Figure~\ref{fig:rateenc}, but rather shift it upwards by two orders of magnitude. For the Solar system, we would then get $\Gamma (\mathrm{R_\odot}) \sim 10^{-4}\;\mathrm{Gyr}^{-1}$, exactly the value given by \citet{BojnordiArbab+Rahvar2021}, and the inner boundary of the GHZ could be placed at roughly $R=1\;\mathrm{kpc}$.

\citet{BojnordiArbab+Rahvar2021} argue that the rate of disruptive stellar encounters is the most dependent on the stellar number density. This should be intuitively understandable, as discussed by \citet{Stojkovic+2019SerAJ}, just by examining Equation~\ref{eq:gamma}: cross-section should be treated as a constant, leaving us with two variables, one of which, relative stellar velocity $\langle v \rangle$, does not vary as much. Naturally, this leads to conclusion that the rate of close encounters very well may be irrelevant, or at least redundant, on larger scales (i.e., for constraining the GHZ) while still applicable on smaller scales (e.g., when considering substructures and local over-densities).

Despite not being of only dynamical significance, the Galactic gas reservoir with its redistribution and its state, should also be considered. In particular, close encounters with gaseous local over-densities such as giant molecular clouds (GMCs), regions of higher star-formation primarily located in spiral arms \citep{Hou+2009}, can have severe effects on the orbits of stars \citep[e.g.][]{Fujimoto+2023}, and moving through GMCs can even be hazardous for the planetary systems orbiting those stars \citep{Kokaia+Davies2019}. Undoubtedly, as previous works suggest, gas can affect the motions of stars but only if local over-densities are present, or rather specific distribution of gas and its flows, typically related to non-axisymmetric features such as bars or spirals \citep[e.g.][]{Berentzen+2007,Yu+2022}. In our axisymmetric galaxy model, we have little to no reason to assume that the inclusion of gas would affect radial migrations or mixing, as the distribution of gas would also be axisymmetric and uniform (unless we specifically model it differently, which is beyond the scope of this paper). However, in a more complex galaxy model, which includes one or more non-axisymmetric features, the possible effects of gas should be modeled or, at least, carefully discussed.

\subsection{Stars evolving from non-circular to circular orbits}

\begin{figure}
\includegraphics[width=\columnwidth,keepaspectratio=true]{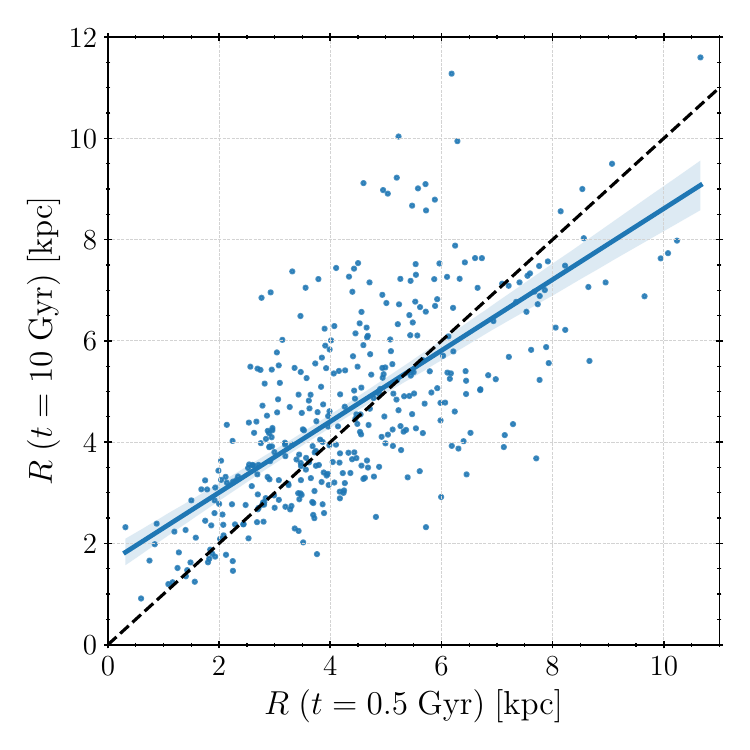}
\caption{Initial versus final galactocentric distance of stars evolving from non-circular to nearly circular orbits, plotted with linear regression line with $95\%$ confidence interval. The dashed black line corresponds to the $y=x$ line.}
\label{fig:evolvetocirc}
\end{figure}

As previously mentioned, stars whose orbital circularity increases are the most interesting from an astrobiological point of view, as these stars evolve toward circular orbits making their environmental conditions stable. Once they evolve to a circular orbit, such stars are likely to permanently reside at a given galactocentric radii making a stable population that can be used to constrain the confinements of the GHZ. We will consider the sub-sample of stars whose circularity increases significantly enough (i.e. from non-circular to nearly circular orbits), regardless of galactocentric distance, since as earlier discussed the GHZ boundaries cannot be effectively constrained with the rate of close encounters. As the previous analysis has shown, these stars (initially in non-circular orbits) exclusively inhabit the inner to central regions of the disk, and this sub-sample represents merely a $3.8\%$ of the stars in the whole sample. The best candidates from this sub-sample should be the stars that stabilize on nearly circular orbits after they have migrated outwards. Thus, we show the initial versus final galactocentric distance of the stars from this sub-sample with a $y=x$ dashed black line to facilitate visual separation of inward from outward migrators in Figure~\ref{fig:evolvetocirc}. Additionally, we show a linear regression line with $95\%$ confidence interval. It is clear that the regression line is not a perfect fit, nor is linear or any other particular functional relationship expected, as the most stars are scattered around. However, it is included for the purpose of an easier assessment of whether there is a global trend and how it deviates from the dashed black line separating migrators.

Interestingly, there are both inward and outward migrators in this sub-sample, but outwards migrations are slightly more prevalent (with net outwards migrations of $11.5\%$) than in the general population ($8\%$). However, at the smaller galactocentric distances, more stars tend to migrate outwards, while it is the opposite case for the larger ones (although there are fewer stars there falling into this sub-sample). This agrees with previously noticed asymmetry in $R-\Delta R$ distribution (Figure~\ref{fig:deltas}), indicative of efficient radial mixing in the region. Median galactocentric distances are $4.168\;\mathrm{kpc}$ and $4.358\;\mathrm{kpc}$ for the initial and the final ones, respectively. Mean values for both are slightly larger than the median, indicating that the distribution is skewed to the right (i.e., higher end). This implies (in a less obvious manner than by examining Figure~\ref{fig:evolvetocirc} for example) that migrations outwards in this sub-sample are, on average, of a higher magnitude.

The most fascinating stars from this sub-sample are the ones scattered above the regression line. These stars migrate farther away before stabilizing in nearly circular orbits in the Solar neighborhood and beyond. Their region of origin can vary a lot, going as low as $R(t=0.5\;\mathrm{Gyr})<3\;\mathrm{kpc}$. The result is particularly interesting given that the observations of the Galactic stellar populations support this, including the outward motion of the Solar System \citep{Wielen+1996}. Naturally, this highlights the need to explore the Solar neighborhood, its stellar population, and the migrations of stars in detail.

\subsection{Broad Solar neighborhood}
\begin{figure}
\includegraphics[width=\columnwidth,keepaspectratio=true]{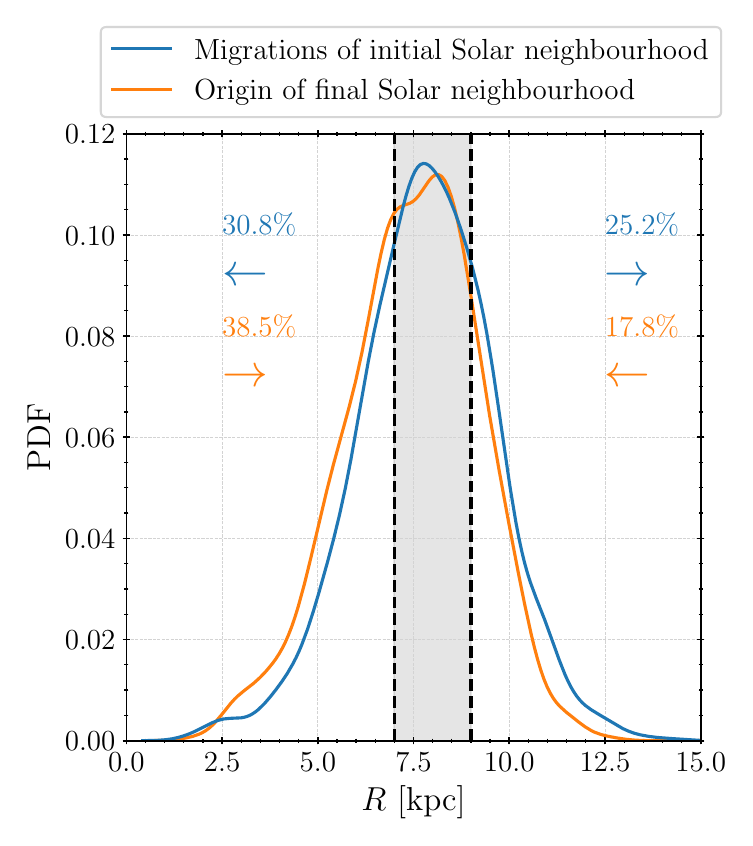}
\caption{Probability density function of galactocentric distance $R$ for stellar populations. The blue line represents the final distribution of the initial stellar population of the broad Solar neighborhood, and the orange one represents the origin of its final stellar population (as indicated by the legend). The Solar neighborhood, as well as percentages and directions of migrators, are clearly indicated.}
\label{fig:solarpdf}
\end{figure}

Given that the presented galaxy model mimics the Milky Way, the traditional, annular GHZ shape can be considered. The conservative estimate from \citet{Lineweaver+2004} place this region between $R=7\;\mathrm{kpc}$ and $R=9\;\mathrm{kpc}$. This region is further referred to as a broad Solar neighborhood and specifically, the orbits of its initial and final stellar population are explored. Figure~\ref{fig:solarpdf} presents the probability density function of galactocentric distance $R$ for stellar populations that, at some point, reside in the broad Solar neighborhood. The blue line is the final distribution of the initial stellar population - the present-day distribution of stars \emph{born} in the Solar neighborhood, i.e., after radial migrations and mixing. The orange line represents the origin of the final stellar population - the distribution of \emph{birth} radii of stars which represent the present-day stable population of the Solar neighborhood, i.e., before migrations and mixing. Since the distributions are not perfectly symmetrical, we list appropriate basic properties (mainly median value, first $Q_1$ and third $Q_3$ quartile, and interquartile range $\mathrm{IQR} = Q_3 - Q_1$) in Table~\ref{tab:solardist}. In addition, we calculate the expected boundaries of these distributions: lower $\mathrm{MIN}= Q_1-1.5\;\mathrm{IQR}$ and upper $\mathrm{MAX}= Q_3+1.5\;\mathrm{IQR}$, thus excluding outliers, i.e., extreme migrators.

\begin{table}
\begin{threeparttable}
\caption{Properties of distributions represented in Figure~\ref{fig:solarpdf}: median (essentially second quartile $Q_2$), first $Q_1$ and third $Q_3$ quartile, and interquartile range $\mathrm{IQR} = Q_3 - Q_1$. The expected boundaries calculated as $\mathrm{MIN}= Q_1-1.5\;\mathrm{IQR}$ and $\mathrm{MAX}= Q_3+1.5\;\mathrm{IQR}$ are also listed.}
\label{tab:solardist}
\begin{tabular}{ccccc|cc}
\toprule
\headrow Population & $Q_2$ & $Q_1$ & $Q_3$ & $\mathrm{IQR}$ & $\mathrm{MIN}$ & $\mathrm{MAX}$ \\
           &    (kpc)     & (kpc)     & (kpc) & (kpc) & (kpc) & (kpc)  \\
\midrule
Initial\tnote{a}        & 7.85       & 6.73     & 9.01 & 2.28 & 3.31 & 12.43 \\ 
\midrule
Final\tnote{b}        & 7.55       & 6.33      & 8.66 & 2.33 & 2.84 & 12.16 \\

\bottomrule
\end{tabular}
\begin{tablenotes}[hang]
\item[a]Migrations of initial Solar neighborhood
\item[b]Origin of final Solar neighborhood
\end{tablenotes}
\end{threeparttable}
\end{table}

While we refer to this assumed GHZ as the broad Solar neighborhood, it should be apparent that the zone is quite narrow when radial migrations are considered. It is also a crowded zone, with a majority of stars moving in and out (thus efficiently mixing in an even broader radial zone), and less than half of the total stellar population inhabits this region at all times. The final stellar population of this zone came primarily from the inner parts of the disk ($38.5\%$ of the population, as opposed to $17.8\%$ that came from outer parts). This is in good agreement with previous studies. The simulations performed by \citet{Roskar+2008} indicate that about half the stars in the Solar neighborhood came primarily from the inner parts of the galaxy. \citet{Hayden+2018}, analyzing the data from the Gaia-ESO Survey on metallicity and kinematics of stars from the Solar neighborhood, also implied that it is likely that stars of higher metallicity have migrated to the Solar neighborhood from the inner parts of the galactic disk. While studying the influence of satellites on radial mixing in the galactic disk, \citet{Bird+2012} mention that $20\%$ of stars that end up in the Solar annulus started at $R<6\;\mathrm{kpc}$. Interestingly, this is similar to the results of this work: $25\%$ of the present-day stable population of the Solar neighborhood originate from smaller galactocentric radii , $R\leq6.33\;\mathrm{kpc}$ (parameter $Q_1$, Table~\ref{tab:solardist}).

The initial stellar population of this zone also migrates out in both directions. It is important to highlight that inward migrations are more prevalent ($30.8\%$) than outward counterparts ($25.2\%$), opposite to the global trends. However, it is expected due to radial mixing, as shown in our previous analysis and observations, in particular because of the asymmetry in $R-\Delta R$ distribution (Figure~\ref{fig:deltas}). Not only do the stars \emph{born} in this region have a higher possibility of inward migrations, but they are also more likely to migrate farther (compared to their outward counterparts). This brings us to the asymmetry of the two distributions around the initially assumed GHZ. Despite the global net positive radial migrations outwards, these distributions do not have a symmetric spread around the broad Solar neighborhood - the spread inwards is of a higher magnitude. If we would assume some initial metallicity gradient, this would imply that the metal-poor stars, that are \emph{born} in the Solar neighborhood, predominantly migrate to the inner regions, and vice versa for the metal-rich stars, which effectively flattens the metallicity gradient \citep[e.g.][]{Roskar+2008,Minchev+Famaey2010,Minchev+2011,Vincenzo+Kobayashi2020}.

Moreover, the zone of efficient radial mixing is much wider than the assumed GHZ, as should be evident from the mid-spread (i.e., $\mathrm{IQR}$ in Table~\ref{tab:solardist}) of two presented distributions, as well as from their expected boundaries. Hence, our results suggest that the region roughly between $R=3\;\mathrm{kpc}$ and $R=12\;\mathrm{kpc}$ represents the zone of radial mixing. Within such a region, it is possible to further constrain the limits of the GHZ based on other (e.g., metallicity) grounds. The stars, which at some point reside in the broad Solar neighborhood, either originate from or migrate to this wider zone, making it challenging to distinguish habitable from non-habitable systems solely on this account if no other additional constraints are applied. Not all stars in stable orbits residing in the GHZ should be considered habitable systems. For example, some systems might come from the outer regions of the Disk and might not be sufficiently metal-rich to form Earth-like planets. Generally, it might be the case that the star was born in a too-metal-rich (or even too-metal-poor) environment, thus a non-friendly one, before migrating to a more stable orbit in the GHZ.

\subsection{On the concept of a Galactic Habitable Zone}

While a typical earth-like planet in a local universe is more likely to be found in a spheroid-dominated galaxy (rather than a disk-dominated one), as \citet{Zackrisson+2016} reported, we will focus closely on previous studies that examined our own Galaxy, since our galaxy model represents the Milky Way. As already mentioned in the introductory section, \citet{Lineweaver+2004} constrained the GHZ to an annular zone between $7\;\mathrm{kpc}$ and $9\;\mathrm{kpc}$ that spreads outwards, while \citet{Prantzos2008} argued that the GHZ annulus is more likely to expand, possibly making the whole disk suitable for hosting life. Interestingly, \citet{Gowanlock+2011} reported that the most habitable regions are located in the inner disk ($\sim2.5\;\mathrm{kpc}$) around the mid-plane (i.e., above or below). On the contrary, \citet{Vukotic+2016} found the outskirts to be the most habitable, constraining the GHZ annulus between $10\;\mathrm{kpc}$ and $15\;\mathrm{kpc}$. \citet{Spinelli+2021} studied the history of the GHZ while considering the effects of gamma-ray bursts and supernovae and found that, indeed, the outskirts are more habitable during early times, but the GHZ shifts inwards, ending as an annulus between $2\;\mathrm{kpc}$ and $8\;\mathrm{kpc}$ at present. While studying the effects of interactions with giant molecular clouds, \citet{Kokaia+Davies2019} gave two estimates of the GHZ annulus for the thin disk (between $5.8\;\mathrm{kpc}$ and $8.7\;\mathrm{kpc}$) and the thick disk (between $4.5\;\mathrm{kpc}$ and $7.7\;\mathrm{kpc}$). Taking only dynamical constraints into account, present results place the whole of the broad Solar neighborhood into the GHZ annulus, extending it toward the inner galactocentric radii -- a birthplace of our own planetary system.

All these different results suggest that habitability estimates are far too sensitive to initial assumptions, fine-tuning of the models, experiment design, and habitability determining factors explored \citep[for a review of different factors influencing galactic habitability, see, e.g.][]{Stojkovic+2019SerAJ}. Considering the effect of stellar migrations on galactic habitability, this work has demonstrated that, in contrast to traditional GHZ studies, the boundaries of such an annulus are not likely to be firmly constrained. However, our results also did not imply that the whole disk is likely to be significantly populated with habitable systems, at least when considering close encounters and the secular evolution of stellar orbits. 

Good agreement of our results, employing the axisymmetric galaxy model with mild levels of dynamical heating, with previous works that include models with transient spiral structure, should not come as a surprise. In particular, \citet{Brunetti+2011} emphasize the importance of taking into account the bar while studying the impact of radial migration on the chemical evolution of the Milky Way. Similarly, \citet{Minchev+2012} mention that the bars are the most effective drivers of radial migration. \citet{Filion+2023} found that the stars in the outer zones likely originated at smaller radii and had their orbits evolve outwards, which has particularly significant implications for habitability. It suggests that, in the presence of bars, the enrichment of the galactic outskirts with potentially habitable stellar systems can happen at a faster pace. However, it should not be considered that the presence of spiral arms is negligible in a dynamical sense. Even from a strictly dynamical perspective, spiral arms are an important driver of radial mixing in the central region, which expands the GHZ (or blurs its boundaries). We also have to acknowledge that they represent local over-densities and influence the rate of close encounters, which would be higher than in the inter-arm regions. Thus, the complexity of robust habitability considerations is evident.

The next generation of numerical GHZ models should effectively consider the motion of the stellar component at the mass resolution of individual stars in order to constrain the parts of the Disk that are populated with stellar systems that are most likely to be habitable. However, running a full-scale version of such numerical models will require a substantial amount of computing resources. We have demonstrated that a test particle approach gives fiducial results and might be used as a path-finder for developing such large-scale projects. Thus, in the meantime, more reliable conclusions will be obtained with advanced models that include the effects of GMCs and quasi-periodical spiral-arm crossings on the galactic stellar orbits, as well as more nuanced habitability considerations. In addition, the effects of the galaxy-dwelling environment should also be accounted for. The presence of satellites and minor mergers predominantly affect the orbits of stars in the outskirts directly \citep{Bird+2012,Carr+2022}. They also cause perturbations that lead to the formation of non-axisymmetric features and drive radial migrations and mixing indirectly. Long-lived non-axisymmetric structures can also form due to quick galaxy flybys \citep[e.g.][]{Pettitt+Wadsley2018,Mitrasinovic+Micic2023}, in which case the influence on the outskirts should be brief unless these interactions happen on inclined orbits and lead to the formation of warps \citep{Kim+2014}. Cosmological and zoom-in simulations are possibly the best tool for studying these environmental effects and their influence on habitability. They are already utilized to explore the habitability of the entire zoo of galaxies \citep{Forgan+2017,Stanway+2018,Stojkovic+2019MNRAS}, many of which display a diversity of shapes and substructures. However, we can also employ isolated N-body or hydrodynamical simulations and other similar methods, essentially utilizing various tools at our disposal. Thus, we obtain richer, more detailed, and more provocative insight into the underlying physics of habitability.

\section{Conclusions}

Advances in studies of galaxy evolution made it clear that stellar radial migrations are commonplace that can explain the observed radial metallicity gradient. We assumed that the radial stellar migrations, largely overlooked (or unknown at the time) in previous habitability considerations, can affect the traditionally constrained GHZ. To check our hypothesis and study this effect, we performed N-body simulations with a realistic model of the Milky Way and artificially added $10^4$ stars whose motions we analyzed.

We confirmed the previous studies \citep[e.g.][]{Roskar+2008,Roskar+2012,Frankel+2020}, finding that there is a small but significant radial shift outwards and that radial migrations are (primarily and predominantly) driven by diffusion in angular momentum while the radial heating is much weaker in comparison (i.e., that churning is much more prevalent than blurring). Most importantly, we demonstrated that radial migrations are not negligible even in a simple axisymmetric model with mild levels of dynamical heating.

The results of our N-body model suggest that, even at sub-Solar galactocentric radii, the rate of close encounters is not large enough to present a hazard for most of the stellar systems with habitable potential. While we argue that the rate of close encounters is not a sufficient condition to constrain the GHZ, its importance is recognized for the local habitability considerations. The net outward diffusion of the stellar component can populate galactic outskirts with stellar systems that have originated in the metal-rich inner parts of the Disk and are thus likely to host Earth-like planets.

Additionally, we have considered stellar systems that evolve from non-circular to circular orbits and found that these systems typically migrate outwards. We have also considered systems with stable nearly-circular galactic orbits in the vicinity of the Solar neighborhood. Such stellar systems are likely to have a stable galactocentric radius making a permanent habitable population at a given annular location. While examining the Solar neighborhood, we have found that the region roughly between $R = 3$ kpc and $R = 12$ kpc represents the zone of efficient radial mixing, which can blur the boundaries of the GHZ. The stellar systems that stabilize on nearly-circular orbits in the Solar neighborhood migrate both from inside and outside parts of the disk. The stars that migrate from the outside should have smaller metallicity and are less likely to host planets. Thus, not all stars on stable orbits in the Solar neighborhood should be considered habitable. With the migration flux from the inner parts being significantly higher, the present-day Solar neighborhood should be more populated with stars that have originated in the inner disk and have higher metallicity -- likely to host planets. As a likely member of such a population, the Solar system can be considered as a typical habitable system in the Milky Way, in contrast to what is often asserted within the framework of the so-called "rare Earth hypothesis".

\begin{acknowledgement}
We thank the anonymous reviewer for insightful comments and suggestions that improved the quality of our paper. The python packages \texttt{matplotlib} \citep{Hunter2007}, \texttt{seaborn} \citep{Waskom2021}, \texttt{numpy} \citep{Harris2020}, \texttt{scipy} \citep{Virtanen2020}, \texttt{pandas} \citep{McKinney2010}, and \texttt{pynbody} \citep{pynbody} were all used in parts of this analysis.\end{acknowledgement}

\paragraph{Funding Statement}

This research was supported by the Ministry of Science, Technological Development and Innovation of the Republic of Serbia (MSTDIRS) through the contract no. 451-03-47/2023-01/200002 made with Astronomical Observatory of Belgrade.

\paragraph{Competing Interests}

None

\paragraph{Data Availability Statement}

The data generated in this study are available from the corresponding author, A.M., upon reasonable request.

%\endnote in some journals will behave like \footnote; and \printendnotes will not output anything. 
\printendnotes

\printbibliography
%\appendix
%\section{Example Appendix Section}

\end{document}